# Magnetization and Anisotropy of Cobalt Ferrite Thin Films

F. Eskandari[1,2], S.B. Porter[1], M. Venkatesan[1], P. Kameli[1,2], K. Rode[1] and J.M.D. Coey[1]

[1]School of Physics and CRANN, Trinity College, Dublin 2. Ireland.
[2]Department of Physics, Isfahan University of Technology, Isfahan 84156–83111, Iran.

Abstract:

The magnetization of thin films of cobalt ferrite frequently falls far below the bulk value of 455 kAm$^{-1}$, which corresponds to an inverse cation distribution in the spinel structure with a significant orbital moment of about 0.6 $\mu_B$ that is associated with the octahedrally-coordinated Co$^{2+}$ ions. The orbital moment is responsible for the magnetostriction and magnetocrystalline anisotropy, and its sensitivity to imposed strain. We have systematically investigated the structure and magnetism of films produced by pulsed-laser deposition on different substrates (TiO$_2$, MgO, MgAl$_2$O$_4$, SrTiO$_3$, LSAT, LaAlO$_3$) and as a function of temperature (500-700°C) and oxygen pressure (10$^{-4}$ – 10 Pa). Magnetization at room-temperature ranges from 60 to 440 kAm$^{-1}$, and uniaxial substrate-induced anisotropy ranges from +220 kJm$^{-3}$ for films on deposited on MgO (100) to -2100 kJm$^{-3}$ for films deposited on MgAl$_2$O$_4$ (100), where the room-temperature anisotropy field reaches 14 T. No rearrangement of high-spin Fe$^{3+}$ and Co$^{2+}$ cations on tetrahedral and octahedral sites can reduce the magnetization below the bulk value, but a switch from Fe$^{3+}$ and Co$^{2+}$ to Fe$^{2+}$ and low-spin Co$^{3+}$ on octahedral sites will reduce the low-temperature magnetization to 120 kAm$^{-1}$, and a consequent reduction of Curie temperature can bring the room-temperature value to near zero. Possible reasons for the appearance of low-spin cobalt in the thin films are discussed.

Keywords; Cobalt ferrite, thin films, pulsed-laser deposition, low-spin Co$^{3+}$, strain engineering of magnetization.

## 1 Introduction

The spinel ferrites are an important family of insulating ferrimagnetic oxides, widely used as soft high-frequency magnetic materials. Their general formula is $MFe_2O_4$ where the iron is ferric $Fe^{3+}$ and M is a divalent transition metal cation, such as $Mg^{2+}$, $Mn^{2+}$, $Fe^{2+}$, $Co^{2+}$, $Ni^{2+}$ or ½($Li^+$+$Fe^{3+}$) [1]. All the ferrites are ferrimagnetic insulating with a high Curie temperature $T_c$, and. with the exception of $Fe_3O_4$ all are insulating. None except $CoFe_2O_4$ exhibits much anisotropy.

The cubic spinel structure, space group $Fd\bar{3}m$ illustrated in Fig. 1, is formed of a cubic close-packed array of oxygen anions slightly displaced from their ideal positions, with the cations occupying tetrahedral $8a$ sites [A-sites], which have cubic $\bar{4}3m$ point symmetry and octahedral $16d$ sites {B-sites}, which have trigonal $\bar{3}m$ point symmetry. There are 56 atoms in the unit cell. The 'normal' cation distribution has the divalent cations on A-sites, but the 'inverse' distribution is more common, where one of the two ferric ions occupies the A-sites, and the other ferric ion and the $M^{2+}$ cations are on B-sites. The ferrite of interest to us here is $CoFe_2O_4$ (CFO), which usually has a near-ideal inverse cation distribution in the bulk

$$[Fe^{3+}]\{Co^{2+}Fe^{3+}\}O_4, \qquad (1)$$

with $Fe^{3+}$ cations occupying the $8a$ sites. The cation distribution can be modified by heat treatment [2], and quenching increases the occupancy of A-sites by $Co^{2+}$. The accepted value of the lattice parameter is 839.2 pm, although the value varies slightly with the sample stoichiometry and preparation method.

$Fe^{3+}$ ($3d^5$; $t_{2g}^3 e_g^2$) has S = 5/2 and a spin moment of 5 $\mu_B$. High-spin $Co^{2+}$ ($3d^7$ $t_{2g}^5 e_g^2$) has S = 3/2 and a spin moment of 3 $\mu_B$, but in B-sites the cobalt can also have a significant unquenched orbital moment of ~ 0.6 $\mu_B$ [3-5], which is responsible for the strong cubic anisotropy $K_1^c \approx 290$ kJm$^{-3}$ with <100> easy directions. Although the moment of an isolated

Co$^{2+}$ ion aligns along a local <111> trigonal axis [3], the resultant bulk anisotropy lies along <100> [6]. The net moment is about 3.6 $\mu_B$ per formula at room temperature (RT), and the magnetization of bulk samples is 455 kAm$^{-1}$ or 86 Am$^2$kg$^{-1}$ (86 emu/g), based on the X-ray density of 5290 kgm$^{-3}$. The ferrimagnetic Néel temperature of CoFe$_2$O$_4$ is 790 K, so the ground state, $T = 0$ values are slightly higher.

Another consequence of the unquenched orbital moment on the Co$^{2+}$ is an exceptionally-large magnetostriction. Originally measured by Bozorth in 1955 [7], the value of $(3/2)\lambda_{100}$ for a Co$_{0.8}$Fe$_{2.2}$O$_4$ crystal was found to be -885 ppm, corresponding to a tetragonality $(a_\parallel - a_\perp)/a$ of almost 1 %. The tetragonality has been measured in unsaturated bulk material by synchrotron X-ray diffraction [8]. A similar value of magnetostricton (-845 ppm) was recorded recently for a crystal of nominal composition CoFe$_2$O$_4$ [9], but, the magnetostriction of a Co-rich crystal, Co$_{1.1}$Fe$_{1.9}$O$_4$, was much less, (-375 ppm) [7], and the values can be quite variable. A result of the magnetostriction is that an imposed strain ε along <100> leads to a uniaxial anisotropy [10]

$$K_u \approx (3/2)\lambda_{100}\varepsilon E \qquad (2)$$

where $E$ is Young's modulus ($C_{11}$), which is 257 GPa [11]. A 1% biaxial compression therefore leads to an easy-plane anisotropy $K_u \approx$ -2 MJm$^{-3}$. In an alternative formulation for biaxially strained cubic films is [12]

$$K_u = (3/2)\lambda_{100} (C_{11} - C_{12}) (a_\parallel - a_\perp)/a \qquad (3)$$

where $C_{12}$ = 106 GPa [11]

The magnetic anisotropy of thin films of CoFe$_2$O$_4$ is exceptionally-sensitive to substrate-induced strain [10, 13-18].

Thin films of $CoFe_2O_4$ with good chemical and mechanical stability have attracted some attention as potential perpendicular recording media [19], as tunnel-barrier spin filters [20-22] due to the spin-dependent bandgap [23], as magneto-optic media [1] and as magnetostrictive films [24].

There have been many reports of preparation of $CoFe_2O_4$ in thin film form using pulsed-laser deposition (PLD) [10, 13-15, 19, 24-31], and the literature also includes reports of films produced by rf sputtering [16, 17, 32-34], MBE [18, 35-37], CVD [38, 39] and ALD [40]. Authors use a variety of substrates, deposition temperatures, oxygen pressures, laser fluence, sample thickness and thermal treatments, and find a wide range of magnetization, anisotropy and hysteresis. A summary of some of the earlier PLD work is provided in Table 1. What is remarkable is that the magnetization found at room-temperature is usually much less than the bulk value, and only comes close to it in a few instances. This is puzzling, because we cannot lower the magnetization by rearranging the $Co^{2+}$ and $Fe^{3+}$ cations on A- and B-sites in a collinear ferrimagnetic structure.

In this work, we have systematically investigated the effects of thin film growth conditions on structural and magnetic properties of CFO thin films deposited by PLD on various substrates, principally with the aim of explaining the anomalously small values of magnetization that are usually found in CFO films, but often ignored by plotting the y-axis of the magnetization curves in reduced units. We discuss our results in terms of the presence of low-spin $Co^{3+}$ ions on B-sites, such as are found in $Co_3O_4$ [41].

Table 1. Reports of magnetization of $CoFe_2O_4$ films prepared by pulsed laser deposition.

| Substrate | Substrate Temperature, $T_s$ (ºC) | Thickness $t$ (nm) | Oxygen Pressure $P$ (Pa) | Magnetization $M$ (kAm$^{-1}$)* | Reference |
|---|---|---|---|---|---|
| MgO (100) | 600<br>800 | 400 | 4 | 335<br>426 | [25] |
| STO//CoCr$_2$O$_4$ | 600/1000p | 140-430 | 0.1 | 380 | [10] |
| MAO | 500/1000p | 65-900 | 0.1 | ? | [26] |
| MgO (100)<br>STO (100) | 700<br>550 | 80 | 10$^{-5}$ | 180 – 220<br>370 | [13] |
| Al$_2$O$_3$ (0001)<br>SiO$_2$ | 800<br>550 | 40-200<br>33 | 0.2 | ?<br>260 | [19] |
| STO (100) | 500-700 | 70 | 0.002-0.01 | ? | [27] |
| MAO(100) | 175-690 | 200-220 | 1.3 (15% ozone) | 450 ( 5K) | [28] |
| MgO(100)<br>STO(100) | 450 | 200 | 1.3 | 300<br>140/480 | [29] |
| Si(100)/SiO$_2$ | 250-600<br>250 | 135 | 2.9<br>0.7-7.0 | 130-270<br>130-220 | [24, 30] |
| MgO(100)<br>STO(100)<br>BTO(100)<br>LAO(100) | ? | 13-100 | 7 | 260<br>390<br>210<br>280 | [14] |
| MgO(100) | 400 | 50-400 | 2 | 100-185 | [15] |
| Pt(111) | 550-750 | 247-290 | 9 | 180-220 | [31] |

* 1 kAm$^{-1}$ is equivalent to 1 emu/cc or 1.26 G     p post annealed

## 2. Experimental Methods

The films were deposited by PLD onto various single-crystal substrates from a sintered $CoFe_2O_4$ target prepared by sol-gel synthesis. A KrF excimer laser (248 nm wavelength with 25 ns pulse width, Lambda Physics) was used to ablate the ceramic target with a laser fluence of about 2 J cm$^{-2}$ and a pulse repetition rate of 10 Hz. The distance between the substrate and target was fixed at 6 cm. Before deposition, the chamber was evacuated to $2 \times 10^{-4}$ Pa and the substrate was heated to 900°C for 1 hour, followed by cooling to the required deposition temperature $T_s$. After deposition, the films were cooled to room temperature at a rate of about 5°C min$^{-1}$ at constant oxygen pressure. They were normally deposited at 600 °C, but some films were deposited at higher or lower temperature (500-700°C). The oxygen pressure was varied in the range $2 \times 10^{-4} - 10$ Pa. In addition, we have deposited films on many other substrates – MgO (100), MgO (110), MgO (111), $SrTiO_3$ (STO) (001), $MgAl_2O_4$ (MAO) (001), $LaAlO_3$ (LAO) (001), $(La,Sr)(Al,Ta)O_3$ (LSAT) (100) and $TiO_2$ (001). The objective was to determine optimum conditions, which would yield films of the highest quality, magnetization and anisotropy. Reflection high-energy electron diffraction (RHEED) was observed during and after film growth. $\theta - 2\theta$ X-ray diffraction (XRD), $\phi$-scans and reciprocal space mapping (RSM) analysis were carried out to check the crystallinity, orientation and strain of the thin films using a Bruker D8 X-ray diffractometer (Cu K$_{\alpha 1}$; $\lambda$ = 154.05 pm). Film thicknesses were determined by small-angle X-ray reflectivity (XRR) and confirmed in some cases by ellipsometry. Morphology of the films was examined using contact-mode atomic force microscopy (AFM), and composition was checked by energy dispersive X-ray analysis (EDX). Most magnetic measurements were made using a 5 T SQUID magnetometer (MPMS 5 XL, Quantum Design) on films mounted in clear plastic straws with magnetic field applied parallel or perpendicular to the film plane. Measurements on selected films were made using a vibrating-sample magnetometer in fields of up to 14 T (PPMS, Quantum Design) when the anisotropy field was very large.

## 3. Experimental Results

We have examined 60 films of cobalt ferrite. Since there are so many experimental variables, we make progress by changing them one at a time — substrate temperature, oxygen pressure, laser fluence, substrate material. Film thickness was generally in the range 30 – 50 nm; we did not investigate ultra-thin films, but we included one 15 nm film on MgO(100) under preferred conditions (600°C, vacuum). Firstly, in order to find the optimum substrate temperature ($T_s$) for CFO thin film growth, two sets of samples were prepared on MgO(100) substrates, one under vacuum ($2\times10^{-4}$ Pa) and the other in an oxygen pressure of 2 Pa. In each case the magnetization was greatest when $T_s \approx 600°$C. The lattice parameter of MgO, which has an ideal cubic close-packed oxygen lattice, is 421 pm or slightly more than half that of CFO (839 pm). The X-ray diffraction patterns of CFO thin films deposited under vacuum on MgO (100) substrates at different $T_s$ of 500 - 700 °C are shown in Fig. 2a,b. The films are all highly-oriented along the [00l] direction, as evidenced by the single CFO (008) peaks visible in Fig 2. Our $\phi$-scans showed four-fold in-plane symmetry, indicating that the film growth on MgO is quasi-epitaxial. The CFO films on MgO are under in-plane tensile stress, so the out-of-plane lattice parameter ($a_\perp = 837$ pm) is smaller than the cubic bulk value. The tetragonality is $\approx$ -0.6 %. In Fig. 2b, it can be seen that the CFO peak width decreases on increasing $T_s$ from 500 to 650°C.

The oxygen pressure inside the chamber during film deposition is another important factor controlling the film structure and magnetization. Therefore, we deposited the CFO films at different oxygen pressures, ranging from vacuum ($2\times10^{-4}$ Pa) to 10 Pa at a fixed substrate temperature of 600°C. All these films grow epitaxially. Fig. 1c shows that the position of the (008) peak shifts to lower angles with decreasing oxygen partial pressure. Thus, the out-of-plane lattice parameter $a_\perp$ and the tetragonality decrease with increasing oxygen partial pressure.

The strain state of the films has been investigated in more detail by reciprocal space mapping (RSM) around the MgO (113) and CFO (226) peaks for samples grown at 600°C in vacuum. The MgO substrates are often twinned, but different in-plane and out-of-plane lattice parameters are seen for CFO, as well as nonuniformity of the out-of-plane parameter across the film thickness for all except the deposition in vacuum. The RSMs of Fig. 3 correspond to CFO films which are fully strained in the plane of the film with $a_\parallel = 842$ pm. The out-of-plane lattice parameter $a_\perp$ is uniform across the film thickness (56 nm) for CFO deposited at 600°C in vacuum (Fig. 3a) and it is equal to 837 pm. There is an increasingly broad distribution of $a_\perp$ with increasing oxygen pressure. The double MgO spots seen for films deposited in 2 or 10 Pa of oxygen are simply due to substrate twins, but the vertical streaking of the CFO (226) reflection indicates a heterogeneous distribution of $a_\perp$ across the thickness. In 2 Pa, the variation is from 832 to 838 pm, whereas for the 10 Pa film of similar thickness, the variation is from 832 to 842 pm. The vertical strain is progressively relaxed in oxygen, but not in vacuum. Figure 3 also shows RHEED patterns of the same three films, which establish good crystal quality. RMS roughness measured by AFM was < 1 nm. The RHEED indicates that the disorder of the surface increases with increasing oxygen pressure. This is suggested by the disappearance of the faint Kikuchi lines and broadening of the diffraction pattern into streaks corresponding to a decrease in long-range order at the surface in the plane of the film as the pressure is increased to 2 Pa, and the collapse of the pattern into spots at 10 Pa, possibly indicating the existence of ordered islands, or the result of transmission through small crystalline structures on the surface.

The composition of selected films has been checked by EDX analysis. The ceramic target was stoichiometric, with an Fe:Co ratio of 1.98(2). Analysis at multiple points on four films deposited on MgO showed that their composition was uniform, but with a slightly different Fe:Co ratio of 1.80(2) indicating that the films are Co-rich.

Figure 4 shows room-temperature magnetization (M-H loops) of CFO thin films deposited on MgO (001) in vacuum at substrate temperatures of 500 and 600°C. The saturation magnetization in both cases is significantly less than the value of 86 $Am^2kg^{-1}$ (455 $kAm^{-1}$) measured for the bulk ceramic target used for the PLD. Plots of $M_s$ vs $T_s$ for films deposited in vacuum and in a 2 Pa pressure of oxygen are shown in Fig. 5. The moments in vacuum are larger, but in both cases, the maximum falls at $T_s = 600°C$, which is why we limit further studies to films produced at this temperature. We find that there is some effect of laser fluence. Increasing the fluence from 2.1 to 2.7 $Jcm^{-2}$ leads to an increase of $M_s$ from 397 $kAm^{-1}$ to 453 $kAm^{-1}$, which is the bulk value. The magnetization of these films on MgO lies perpendicular to the film plane, with coercivity of up to 400 $kAm^{-1}$ for the high-moment films. Coercivity in the low-moment films is greater. The magnetization curve of the 453 $kAm^{-1}$ sample, which we can regard as a benchmark, is included in Figure 8. There is no hysteresis when the field is applied in-plane, but four-fold anisotropy is associated with the tetragonal symmetry of the film. The intrinsic perpendicular anisotropy $K_u$ can be estimated from the equation

$$K_u = K_\perp - K_s \qquad (4)$$

where the parameter $K_\perp$ is deduced from the area between the perpendicular and parallel magnetization curves, and the shape anisotropy $K_s$ is $-½μ_0 M_s^2$ The anisotropy energy associated with each term is of the form $E = K\sin^2θ$, where θ is the angle between the direction of uniform magnetization and the film normal. From the data in Fig. 4, $M_s$ = 397 kAm$^{-1}$ so $K_s$ = - 99 kJm$^{-3}$. The measured value of $K_\perp$ is 127 kJm$^{-3}$, hence it follows that $K_u$ = 226 kJm$^{-3}$. Substantially larger values of $K_u$ have been reported by Niizeki for samples with very similar magnetization curves using torque measurements [42] or simple estimates of the anisotropy field $\mu_0 H_a = 2K_u/M_s$ [16]. Our estimates of $K_u$ from magnetization curves that exhibit hysteresis can underestimate $K_u$ by up to 35% because of effects of magnetic viscosity in hysteretic easy direction.

Figure 6 shows magnetization curves for samples deposited at 600°C in two different oxygen pressures. The oxygen decreases the magnetization, as summarized in Fig. 7. The magnetization of the sample grown in an oxygen pressure of 10 Pa is only 60 kAm$^{-1}$. This is partly because the Curie temperature of the sample is rather low. The oxygen pressure dependence is opposite to that found in NiFe$_2$O$_4$ and YIG [43, 44]. From the temperature-dependence of $M_s$, $T_c$ is estimated to be about 450 K. The magnetization of these low-moment films on MgO is perpendicular to the plane, and the coercivity becomes very large at low temperature, exceeding 5 T at 100 K. The 15nm film prepared at 600°C in vacuum is different to all the other, thicker films, insofar as the moment lies in-plane. Its magnetization is 389 kAm$^{-1}$.

Next, we turn to results for the other substrates. Films deposited at 600°C in vacuum on the other cuts of MgO, (111) or (110), show lower magnetization than on (100), 280 kAm$^{-1}$ and 254 kAm$^{-1}$, respectively, with less pronounced anisotropy. The results for other substrates, shown in Figure 8, are more interesting. Films on TiO$_2$ (100) and LAO (100) exhibited magnetization of 444 kAm$^{-1}$ and 412 kAm$^{-1}$ respectively, close to the bulk value, but these substrates have a large lattice mismatch of ±10% so there is no epitaxy. The films are polycrystalline. Magnetization lies in-plane and the anisotropy is due to shape, with negligible $K_u$ contribution from substrate-induced lattice strain.

The films on spinel, $MgAl_2O_4$ substrates, are subject to 3.7% lattice mismatch, leading to strong perpendicular expansion. The RSM illustrated in Fig. 9 shows that the in-plane lattice parameter does not follow the template provided by the spinel lattice, but there is nonetheless a broad, correlated distribution of lattice parameters centred at $a_\parallel$ = 831 pm and $a_\perp$ = 850 pm, corresponding to an in-plane compression of 1%, and a similar perpendicular expansion. The tetragonality is 2%. Magnetization curves plotted in Fig. 8 show a magnetization of CFO on MAO of 315 kAm$^{-1}$, with a coercivity of 380 kAm$^{-1}$ and an enormous in-plane anisotropy estimated from the area between the curves as $K_u$ = -1.12 MJm$^{-3}$, far exceeding the shape anisotropy $K_s$ = - 62 kJm$^{-3}$. The anisotropy field $\mu_0 H_a$ is 14 T, and the anisotropy energy, estimated simply as $-½\mu_0 H_a M_s$ is -2.2 MJm$^{-3}$. Both $M_s$ and $K_u$ are reduced in magnitude in a 2 Pa oxygen atmosphere, to 262 kAm$^{-1}$ and -0.58 MJm$^{-3}$, respectively. Films deposited on LSAT behave similarly, with strong negative $K_u$ = -0.65 MJm$^{-3}$ for films made in vacuum.

STO is different. Here the film is oriented, but the magnetization is only 40% of the bulk value. The perpendicular lattice parameter expands a little, but the net anisotropy is perpendicular to the plane, as for MgO, but very weak. The effect of substrate strain is 20 – 40 times less than for MAO or LSAT, and of opposite sign. The anisotropy and magnetization of the films deposited on different substrates is summarized in Table 2.

**3 Discussion.**

3.1 Anisotropy.

We confirm the idea, originally proposed for films on MgO and MAO [10, 36] and confirmed by electronic structure and crystal field calculations [23, 45-49]that the magnetic anisotropy in cobalt ferrite films is largely governed by substrate-induced strain. All our films show 001 oriented growth, except those deposited on TiO$_2$ (001) or LaAlO$_3$ (100) substrates, which have in-plane lattice parameters that are completely mismatched to that of CoFe$_2$O$_4$, being about 10% bigger or smaller. In these two cases, there is effectively no substrate strain, and shape anisotropy dominates (Table 2). The magnetization of these unconstrained films is close to that of bulk CFO. They exhibit easy-plane magnetization with $K_s \sim 100$ kJm$^{-3}$ and isotropic coecivity, $H_c \approx 120$ kAm$^{-1}$.

Table 2. Anisotropy of cobalt ferrite films deposited at 600°C in vacuum on different 100 oriented substrates.

| Substrate | $a$ Parameter (pm) | Growth | $M_s$ (kAm$^{-1}$) | $K_s$ (kJm$^{-3}$) | $K_u$ (kJm$^{-3}$) | Anisotropy |
|---|---|---|---|---|---|---|
| TiO$_2$ | 919 | Unconstrained | 413 | -107 | ~ 0 | Shape in-plane |
| MgO | 842 | Quasi-epitaxial | 397 | -99 | 226 | Strain perpendicular |
| MgAl$_2$O$_4$ | 808 | Oriented, strained | 317 | -63 | -1120 | Strain in-plane |
| SrTiO$_3$ | 782 | Oriented | 183 | -21 | ~ 30 | Strain, weakly perpendicular |
| LSAT | 775 | Oriented, strained | 270 | -46 | -650 | Strain in-plane |
| LaAlO$_3$ | 758 | Unconstrained | 444 | -123 | ~ 0 | Shape In-plane |

The blue line separates the parameters that are greater than or less than that of CFO, 839 pm.

Next come the MAO and LSAT substrates where there is still a large lattice mismatch, but there is now clear 001 axis orientation of CFO, with the wide spread of correlated $a_{//}$ and $a_\perp$ parameters, shown in Fig. 9. The CFO is significantly strained, and according to Eq.3 the tetragonality of 2.2% should lead to uniaxial hard-axis anisotropy of -2.9 MJm$^{-3}$ on account of the cobalt magnetostriction, However, the magnitude of $\lambda_{100}$ in CFO is often much less than the frequently-cited 590 ppm [7], and the resulting $K_u$ will be correspondingly reduced. For example, a value $\lambda_s$ = 225 ppm reported for polycrystalline CFO [50] corresponds to $\lambda_{100}$ = 256 ppm. The tetragonality of CFO on LSAT is 1.9%.

The origin of $\lambda_{100}$ is the high-spin $Co^{2+}$ on B-sites, so we can expect that there are significant variations in the amount of this ion in different films, which provides a link between anisotropy and magnetization. High-spin $Co^{2+}$ on A-sites does not contribute to the anisotropy because there is no orbital moment in cubic $\bar{4}3m$ point symmetry. In DFT calculations, cation distributions were found to be sensitive to epitaxial strain [51]. Furthermore, a study of the influence of substrate strain on $K_u$ in $CoCr_2O_4$, a spinel with the normal cation distribution, reveals that its sign is opposite to that for $CoFe_2O_4$ [52].

Magnetoelastic strain-related anisotropy of CFO on MAO has been discussed previously [34], including a study of 5 nm MBE films which were found to grown epitaxially on MAO [36], but the extremely large easy-plane anisotropy we have observed on spinel is unprecedented; the anisotropy field is more than double anything reported previously [36, 53]. That for CFO on LSAT is also very large. The magnetization of CFO on both substrates is about 2/3 of the bulk value.

STO is a different story. Here the perpendicular 004 and 008 X-ray reflections are broad, but the perpendicular lattice parameter of 841.8 pm indicates a dilation of just 0.3%. Furthermore the magnetization is much reduced, to 40% of the bulk value. The shape anisotropy is therefore weak, $K_s$ = -21 kJm$^{-3}$, and it appears from the magnetization curves in Fig 8 that the uniaxial anisotropy $K_u$ is of similar magnitude but opposite sign, leading to a feeble perpendicular anisotropy $K_\perp > 0$ for the film. The anisotropy mechanism based on the magnetostriction of B-site $Co^{2+}$ does not apply in this case. We think the small moment signals a cation distribution that is quite different to that of bulk CFO.

The films grown on MgO at 600°C in vacuum are of good epitaxial character according to the RSM of Fig 3a, and the RHEED pattern of Fig. 3d which shows traces of Kikuchi fringes. RMS surface roughness for the 56 nm thick film is 0.7 nm. The films have 85 % or more of the bulk magnetization, and the tetragonality of -0.6% corresponds to a positive $K_u$ of 800 kJm$^{-3}$ according to eq. 3, significantly more than our measured value of $K_u$ of 226 kJm$^{-3}$ (Table 2). The discrepancy may arise because the value of $(3/2)\lambda_{100}$ in our films is reduced to about 250 ppm, or because our method of evaluating $K_u$ from the magnetization curves underestimates the value. Torque measurements generally give greater values [16, 34]. Figure 12 summarizes the relation between tetragonality or perpendicular strain and anisotropy $K_u$ in the films.

We have not performed a systematic study of anisotropy vs film thickness for CFO films on MgO, but we found that a 15 nm film was easy-plane, whereas films with $30 \leq t \leq 70$ nm are all easy-axis. A low-moment 5 nm MBE film was also found to be easy-plane [36]. Easy-axis anisotropy persists up to 300 nm, but the thickest PLD films become easy-plane again as the strain is eventually relaxed, and the structure becomes cubic [54]. We should therefore include another term $\kappa_i/t$ in the anisotropy expression (4) to take account of the interface anisotropy; $K_\perp = K_u - K_s + \kappa_i/t$. From the first spin reorientation thickness, we estimate that $\kappa_i$ is easy-plane and approximately - 6 mJm$^{-2}$. Figure 10 sketches the evolution of the anisotropy of CFO films on MgO, marking the dominant anisotropy term in each of three thickness regimes.

3.2 Magnetization.

A remarkable feature of the cobalt ferrite films produced by PLD and other methods is the magnetization, which frequently falls short of the bulk value and is remarkably sensitive to sample preparation conditions. A well-known consequence of thermal treatment of bulk ferrites is to induce deviations from the ideal inverse cation distribution [$Fe^{3+}$]{$Fe^{3+}Co^{2+}$}$O_4$ by moving some $Co^{2+}$ ions over to A-sites, and an equal number of $Fe^{3+}$ ions onto B-sites. The formula is then

$$[Fe^{3+}_{1-x}Co_x]\{Fe^{3+}_{1+x}Co^{2+}_{1-x}\}O_4 \qquad (5)$$

and the moment is thereby increased from $3.5\mu_B$ per formula unit to $3.5(1 + x)\mu_B$, assuming an A-site Co moment of $3\mu_B$. In any case, no permutation of these cations can *reduce* the net magnetization of a collinear ferrimagnetic spin structure. The slight excess of cobalt detected in the EDX analysis of the films, which was independent of the preparation method, does reduce the magnetization a bit. The inverse formula is then $[Fe^{3+}]\{Fe^{3+}_{1-y}Co^{2+}_{1+y}\}O_4$, where an Fe/Co ratio of 1.8 corresponds to $y = 0.07$. The magnetization falls to 80% of the bulk value, but permuting the cations only increases the value.

How then can we get the low values of room-temperature magnetization seen in Table 2 and Figs 5,7. The average of $M_s$ for 35 different CFO films was 240 kAm$^{-1}$, just 53% of the bulk value. There are several possibilities to consider, namely low magnetic ordering temperatures, noncollinear magnetic structures including the effects of antiphase boundaries, low-spin cobalt and oxygen stoichiometry

*Magnetic ordering temperature*. The great majority of our magnetization measurements were made at room-temperature. The reason for not measuring systematically at low temperature is the Curie law paramagnetism of iron impurities in the MgO substrates used for more than half the depositions. The effect is small and linear above 100 K, but increasingly important and nonlinear below. Since $T_c$ of CFO is 790 K, thermal effects at RT/$T_c \approx 0.37$ are expected to reduce the magnetization by about 5% [55]. We checked the Curie temperature of the 600ºC sample with the smallest magnetization, prepared in 10 Pa oxygen pressure (Fig 7), which was estimated to be roughly 460 K, from magnetization measured at temperatures up to 380 K. For all other samples with reduced moment, such as those produced in an oxygen pressure of 2 Pa, $T_c$ is higher and finite temperature effects have little impact on the magnetization at room-temperature.

*Noncollinear magnetic structures.* An explanation commonly advanced for the low moments of spinel ferrite films deposited on MgO is that antiphase boundaries [17, 36, 37, 56] are incorporated into CFO during film growth. Originally proposed to explain the low magnetization and slow approach to saturation of magnetite [57, 58], an antiphase boundary forms where two crystallites that have nucleated independently on MgO (100) grow together. There is therefore a plane of antiferromagnetic B − O − B interactions at the interface which are hard to overcome when the crystallites themselves are of nanoscale dimensions. These effects may well be present in our films grown at low temperatures, but the observation of near-bulk values of magnetization for samples grown at 600°C in vacuum (Fig 8a) and the high crystalline quality of these films (Fig 3a) suggest that antiphase bondaries may be important only in films grown at low $T_s$. These antiphase boundaries are not expected in films on MAO which has the same spinel structure and crystal symmetry as CFO, although there is another type associated with misfit dislocations that has been identified in $Fe_3O_4$ films on MAO [59].

Another possible reason for a reduced moment in an inverse spinel would be a noncollinear or canted spin structure of the B-site cations, which form the majority sublattice. The spins in zinc ferrite, for example, freeze in a random arrangement below about 15 K, but $ZnFe_2O_4$ is a normal spinel, with nonmagnetic $Zn^{2+}$ cations on the A-sites. The presence of A-site $Fe^{3+}$ in inverse spinels ensures that the antiferromagnetic 135º A − O − B superexchange interactions are strong, and determine the collinear Néel state [60].

The next and most plausible explanation for the reduced moment relates to the spin state of the cobalt. In the bulk, B-sites are populated by $Fe^{3+}$ and high-spin $Co^{2+}$, which is stabilized by the trigonal crystal field at the distorted octahedral sites, which have $\bar{3}$m symmetry on account of the deviation of the oxygen 32$e$ site special position parameter $u$ from 3/8. One possibility is that substrate-induced strain switches the $Co^{2+}$ on B-sites to a low-spin state, $3d^7$ $t_{2g}^6 e_g^1$ with S = ½ and a spin moment of 1$\mu_B$. The Curie temperature may be similar, but the magnetization will be much lower, ~ 130 kAm$^{-1}$ or 24 Am$^2$kg$^{-1}$. Some admixture with the normal spinel configuration $[Co^{2+}]\{Fe^{3+}_2\}O_4$ is possible, but as usual it will increase the moment. Cation size, crystal-field stabilization energy and temperature are all factors that determine the spin state, but $Co^{3+}$ is often low-spin on octahedral sites, where it adopts a stable $3d^6$ $t_{2g}^6$ configuration with no moment, S = 0 [61]. The normal cobalt spinel $Co_3O_4$ has nonmagnetic low-spin $Co^{3+}$ on B-sites [41, 62]. Another possibility is therefore the replacement of $Fe^{3+}$ and high-spin $Co^{2+}$ on B-sites by $Fe^{2+}$ and low-spin $Co^{3+}$. This also has the effect of reducing the net moment per formula to 1 $\mu_B$ but it makes A-site $Fe^{3+}$ the majority sublattice. Figure 11 illustrates the spin states of Co ions on the two sites in a one-electron energy level picture. The splitting of the $t_{2g}$ triplet on octahedral B-sites corresponds to compression of the octahedra along the local <111> axis [3]. Table 3 compares the ionic radii and the spin moments in tetrahedral and octahedral sites. It can be seen there that the mean B-site radius of 70 pm in the inverse structure is reduced to 65 pm or 66.5 pm in with the first or second of these cobalt low-spin hypotheses, marked in blue or green on the table, respectively. Any substrate which exerts imperfectly-relaxed compressive strain, those below the blue line in Table 2, may be expected to favour some low-spin cobalt, which can account for the reduced moments observed on MAO, STO and LSAT, and on MgO prepared in oxygen.

The correlation of the low moment observed in CFO films with low-spin cobalt means that strain-induced anisotropy $K_u$ is correspondingly reduced. Low-spin cobalt can provide an explanation for the anomalous behaviour of the films on STO. They have a low moment of 185 kAm$^{-1}$ and a greatly reduced anisotropy with a positive sign (see Table 2). Their low moment suggests conversion of much of the cobalt to a low-spin state, which will reduce the strain-induced anisotropy, and may even change its sign [16, 45, 48]. Furthermore, depositing films on substrates like STO or MgO with an undeformed cubic oxygen sublattice may modify oxygen special position parameter *u*, hence the trigonal distortion of the oxygen around the B-sites, thereby reducing the orbital moment and the magnetostriction $\lambda_{100}$ that controls $K_u$.

Table 3. Ionic radii (in pm) and spin states of Fe and Co cations on in tetrahedral or octahedral oxygen sites

|  | $Fe^{2+}$ | $Fe^{3+}$ | $Co^{2+}$(Hs) | $Co^{3+}$(Hs) | $Co^{2+}$(Ls) | $Co^{3+}$(Ls) |
|---|---|---|---|---|---|---|
| Tetrahedral Ionic radius | 58 | 49/49/49 | 56 | 47 | 49 | 45 |
| spin | 2 | 5/2 | 3/2 | 2 | 3/2 | 1 |
| Octahedral Ionic radius | 78 | 65/65 | 75 | 61 | 65 | 55 |
| spin | 2 | 5/2 | 3/2 | 2 | 1/2 | 0 |

Finally, we consider the effect of oxygen pressure, which strongly reduces the moments of films on MgO (Fig 7) and MAO. The spinel lattice is known to accommodate excess oxygen by means of B-site cation vacancies, oxidizing divalent cations there to the trivalent state. The textbook example is $\gamma Fe_2O_3$, which is really the cation-deficient spinel

$$[Fe^{3+}]\{Fe^{3+}_{5/3} \square_{1/3}\}O_4, \qquad (6)$$

where $\square$ is a vacant B-site. A similar effect in CFO leads to a formula $[Fe^{3+}]\{Fe^{3+}_{7/9} Co^{3+}_{8/9}$ 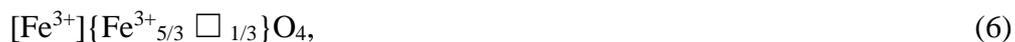 $\square_{1/3}\}O_4$. If the $Co^{3+}$ is low-spin, the ferrimagnetic net moment is 1.1 $\mu_B$ and the A-sites form

the majority sublattice. Furthermore, nonmagnetic low-spin $Co^{3+}$ means that 11/18 of the B-sites are nonmagnetic, which will reduce the magnetic ordering temperature to 350 K or less. We can therefore expect oxygen pressure to reduce both $M_s$ and $T_c$, although admittedly Eq 6 represents an extreme case.

The plot in Fig. 12 shows $K_u$ vs tetragonality. The films prepared in vacuum with $M_s$ > 300 kAm$^{-1}$ show roughly a linear relation. These films have mostly high spin $Co^{2+}$ on B-sites, which provides the strain-sensitive anisotropy. However, the two films with a low moment $M_s$ < 200 kAm$^{-1}$ (MgO at 2 Pa, STO) are different, we think because of the important fraction of low-spin $Co^{3+}$ in these materials.

## 4 Conclusions.

Our systematic broad-brush examination and analysis of cobalt ferrite thin films has revealed a great variety of magnetic properties, often quite different to those of the bulk. The influence of the substrate is evidently critical, as are the deposition temperature and oxygen pressure. Strain engineering is especially rich in CFO, because it works two ways. First is to induce tetragonality in the films, which deforms the strong {100} cubic anisotropy $K_1^c$ associated high-spin $Co^{2+}$ in octahedral sites to give a uniaxial anisotropy $K_u$, which may be easy-axis or easy plane, depending on the substrate. Second, a triaxial compressive strain can convert cobalt from a high-spin to a low-spin state, thereby modifying the magnetization, $M_s$, shape anisotropy $K_s$ and magnetocrystalline anisotropy $K_1^c$. There is therefore a new opportunity to modify anisotropy with strain. The example of CFO on STO shows the cobalt ferrite teetering on an edge, where a slight strain-induced anisotropy change could have a big effect. This suggests suggests a possibility of using piezoelectric substrates to switch, based on cobalt spin-state transitions and the related anisotropy

The next step should be to look carefully for the effects postulated using methods that exhibit atomic-scale sensitivity. These include electron microscopy with atomic scale resolution to evaluate the substrate-induced strain profiles and extended lattice defects, as well as spectroscopic methods such as XMCD to determine the cobalt spin and orbital moments,. Care will be needed to distinguish near-surface ions which can be subject to a reduced crystal field from those deeper in the film where the full crystal field will induce a low spin state. $Co_3O_4$ thin films will be useful reference samples here. Ferromagnetic resonance will be useful to evaluate the anisotropy fields and magnetizations independently.

Cobalt ferrite illustrates the many ways in which a thin film of an oxide containing $Co^{2+}$ can differ from bulk material, and it offers new opportunities for strain engineering of magnetic layers in oxide electronics.

**Acknowledgements.** This work was supported by Science Foundation Ireland as part of the AMBER Research Centre grant SFI/12/RC/2278, and the Centre-to-Centre grant SFI/16/US-C2C/3287. This project has received funding from the European Union's Horizon 2020 research and innovation programme under grant agreement No 737038.

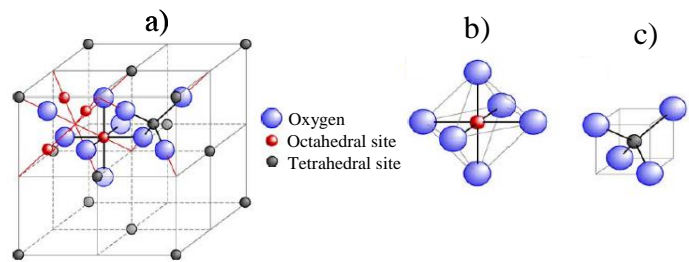

**Fig. 1.** The spinel structure, illustrating the tetrahedral A-sites (black) and octahedral B-sites (red)

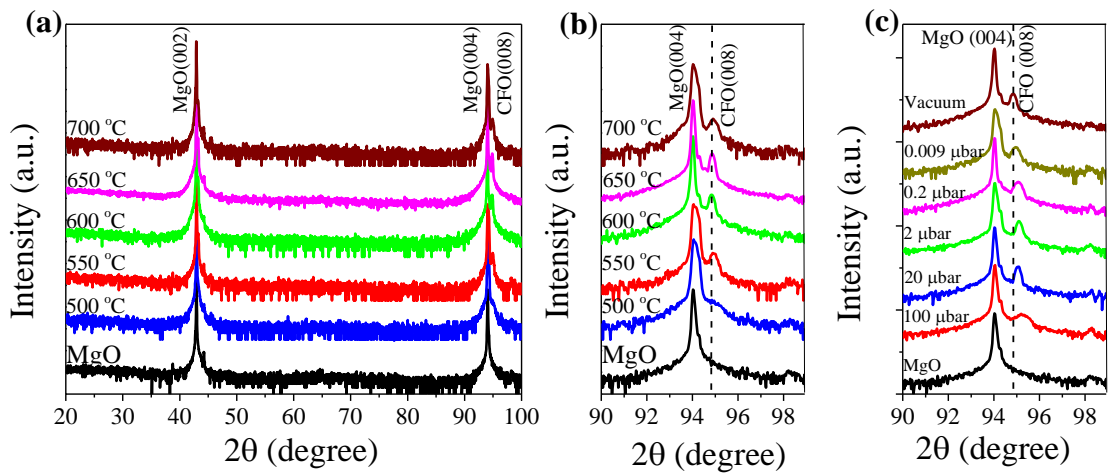

**Fig. 2.** (a), (b) XRD patterns of CFO films on MgO(100) prepared at different substrate temperatures $T_d$ in vacuum, and (c) at different pressures at $T_d$ = 600 ºC.

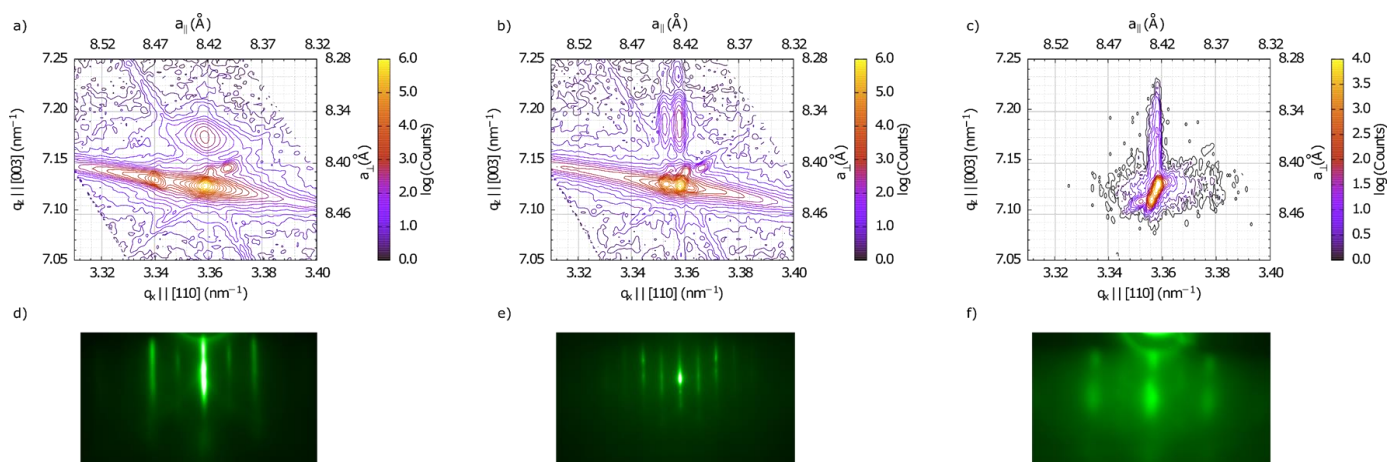

Fig. 3. Reciprocal space maps about the (113) reflection showing in-plane and out of plane lattice parameters for CFO films deposited on MgO (001) at 600°C in (a) vacuum (a) 2 Pa c) 10 Pa and the respective RHEED patterns d), e), f).

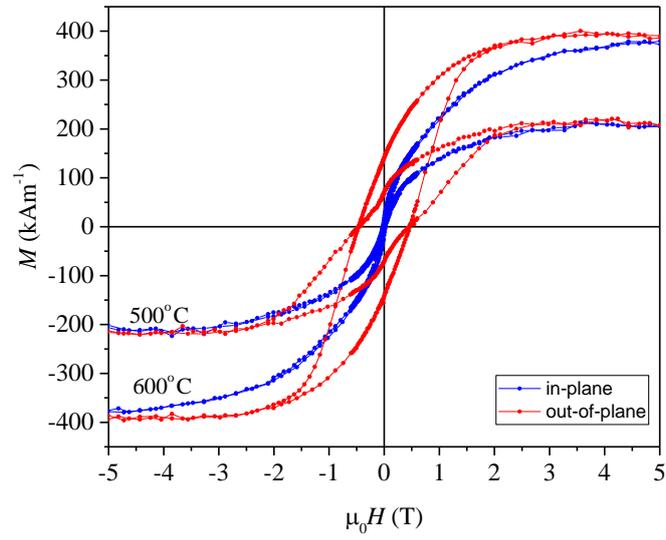

**Fig.4. Magnetization of CFO films deposited on MgO (001) in vacuum at 500 and 600 ºC.**

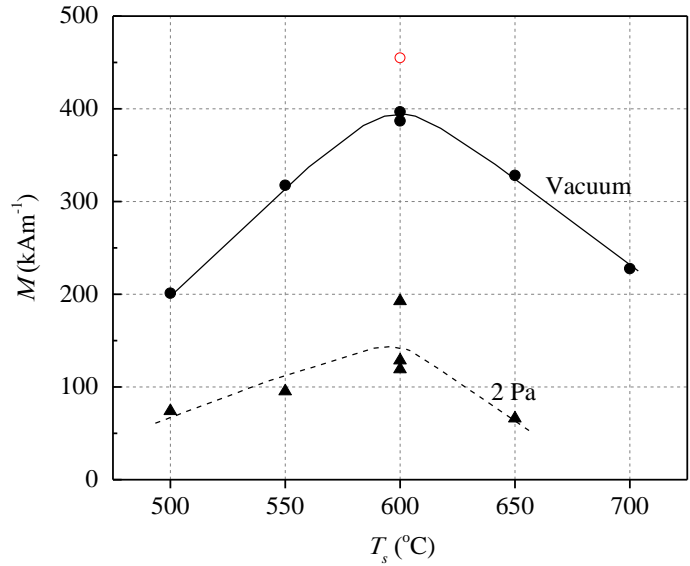

**Fig.5. Magnetization as a function of substrate temperature for CFO films deposited on MgO (001) in vacuum (solid circles), and in 2 Pa solid squares. Laser fluence was 2.1 Jcm$^{-2}$, except for open circle (2.7 J Jcm$^{-2}$).**

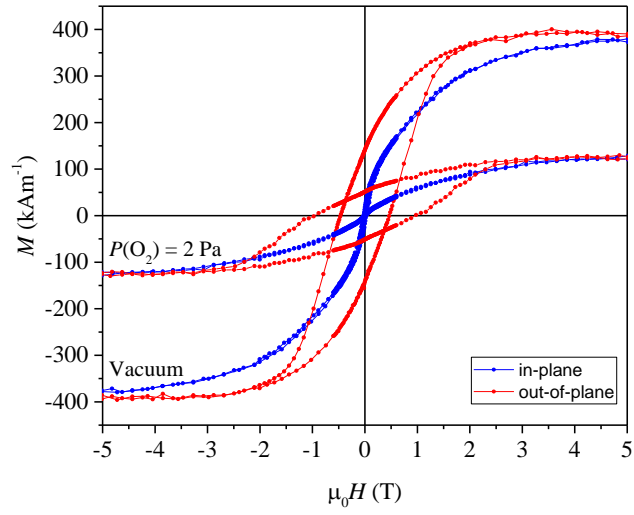

**Fig.6. Magnetization of CFO films deposited on MgO (001) at 600 ºC in 2 Pa and vacuum. The laser fluence was 1.8 Jcm$^{-2}$.**

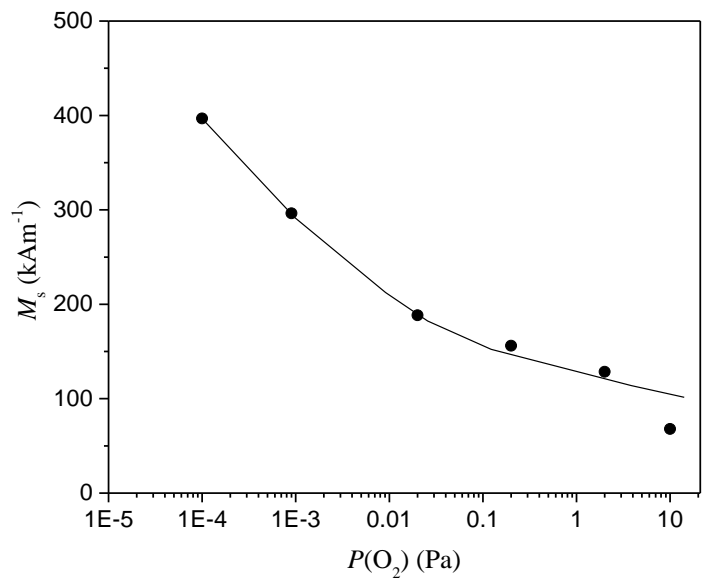

**Fig. 7.** Room temperature magnetization for CFO films deposited on MgO (001) at 600 ºC and different oxygen pressures. Laser fluence was 2.1 Jcm$^{-2}$.

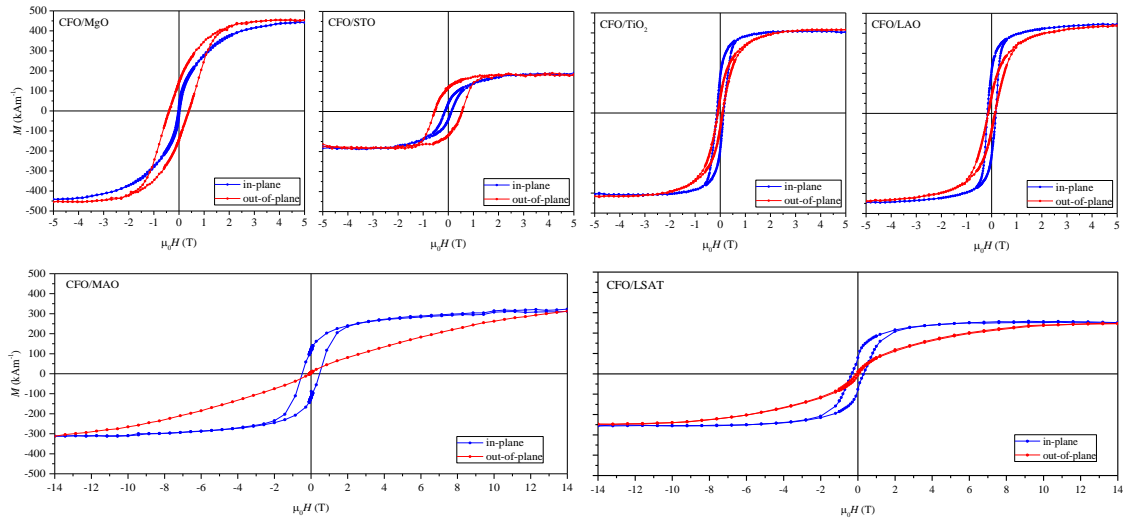

**Fig. 8.** Room temperature magnetization of CFO films deposited on six different (100) substrates at 600°C in vacuum. The graphs are plotted on the same vertical scale and expanded horizontal scales to emphasize the influence of the different substrates.

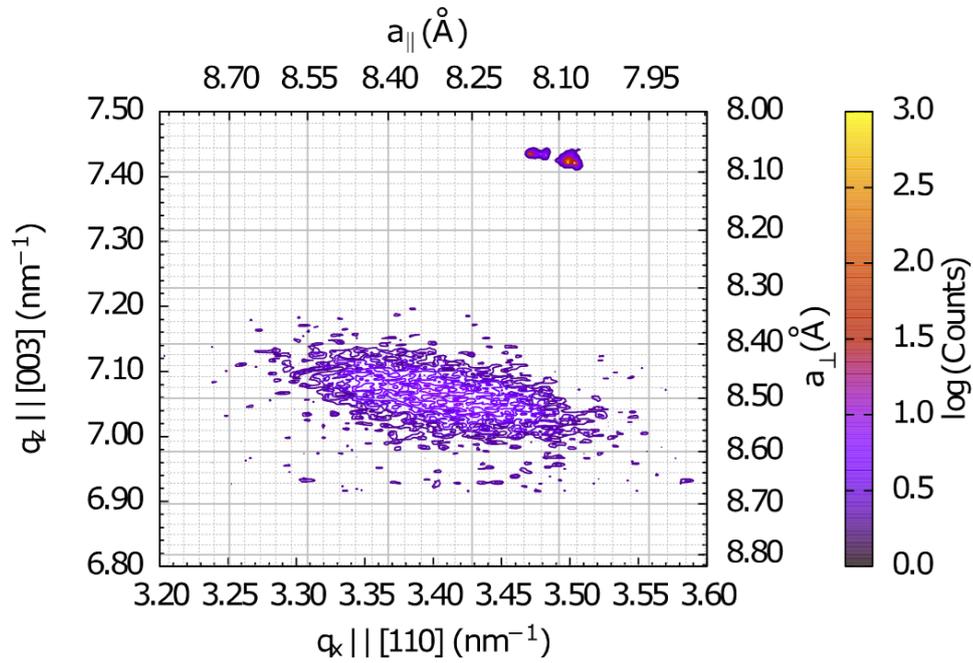

**Figure 9 The reciprocal space map of CFO on MgAl$_2$O$_4$ (100) grown under optimised conditions. MgAl$_2$O$_4$ imposes a strong in-plane compression on the CoFe$_2$O$_4$ film, which causes a dramatic elongation of the out-of-plane lattice parameter to conserve the volume of the unit cell.**

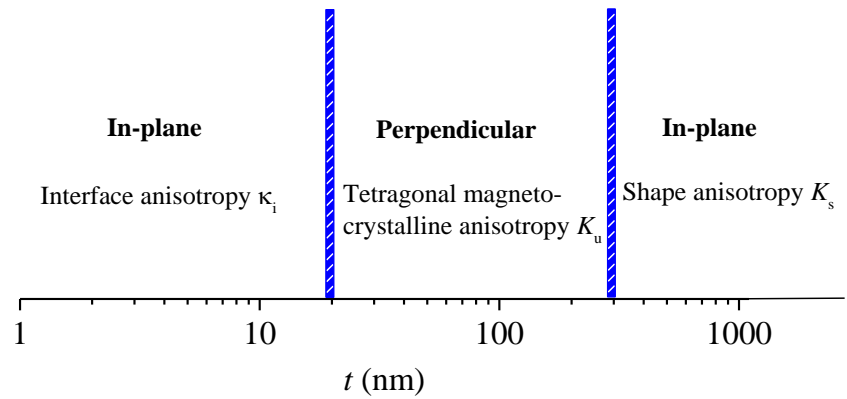

**Figure 10. Magnetization direction for CFO films grown on MgO (100) as a function of film thickness. In the thinnest films, interface anisotropy is dominant, but in the thickest ones, strain relaxation eliminates the tetragonal magnetocrystalline anisotropy, and shape anisotropy then predominates.**

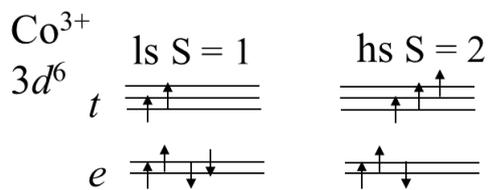
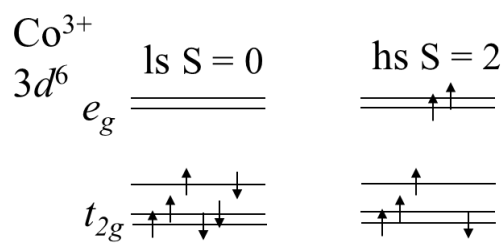

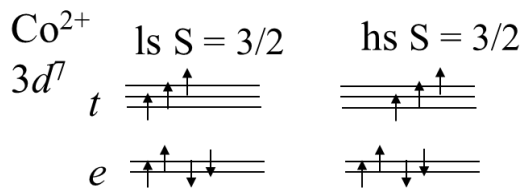
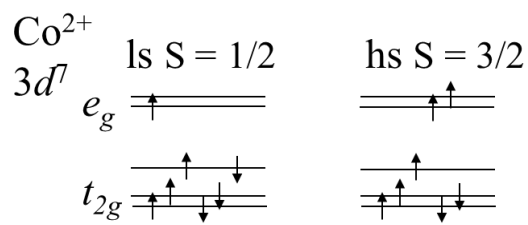

**Figure 11.** One electron energy diagrams for $Co^{2+}$ and $Co^{3+}$ on tetrahedral or octahedral sites in high and low spin states.

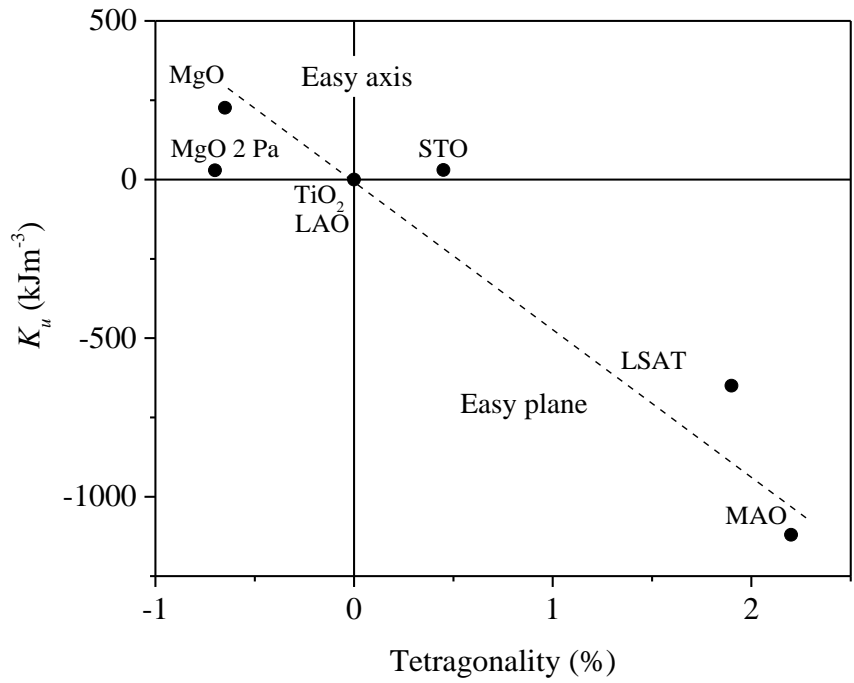

**Figure 12. Plot of uniaxial anisotropy vs tetragonality for CFO films on different substrates.**